\documentclass[10pt,tightenlines, amsmath, amssymb, nofootinbib, prd,
superscriptaddress, showpacs, preprintnumbers]{revtex4}
\usepackage[english]{babel}
\usepackage{graphicx}
\usepackage{mathtext}
\usepackage{indentfirst}
\usepackage{epsfig,amsmath,amsfonts}

\begin{document}

\centerline{\Large \bf Chiral Vortical Effect in Fermi Liquid}

\vspace{10mm}

\centerline{Z.V.  Khaidukov, V. P. Kirilin, A.V. Sadofyev}

\vspace{5mm}

\centerline{Institute for Theoretical and Experimental Physics, Moscow}
\date{}


\begin{center}{\bf Abstract}\end{center}
In this note we consider rotating fermi liquid in the presence of Berry curvature. We argue  that there appears an analogue of chiral vortical effect in the liquid if Berry curvature has a non-vanishing flux through sheets of Fermi surfaces and corresponding chemical potentials are different. We discuss correspondence between relativistic and non-relativistic dispersion type  in vicinity of degeneracy points. We also claim that quantum anomalies in condensed matter context provide a theoretical and experimental testing ground for the origin of chiral effects, their carriers etc.

\section{Introduction}
There has been a lot of interest in chiral effects in literature recently \cite{Kharzeev, SonSurowka, Zakharov, Sadofyev, Zhitnitsky, Volovik2}. Various approaches have been successfully employed to describe these effects, however their physical origin is clearly the existence of chiral anomaly. We will concentrate our attention on the most widely discussed chiral magnetic effect (CME) and chiral vortical effect (CVE) in a vector current of the particle number:

\begin{eqnarray}
\label{chiral}
\overrightarrow{j}_{CME}=\frac{\mu_5}{2\pi^2}\overrightarrow{B}~~,~~\overrightarrow{j}_{CVE}=\frac{\mu_5\mu}{\pi^2}\overrightarrow{\omega},
\end{eqnarray}
where coefficients in front of magnetic field $\overrightarrow{B}$ or vorticity $\overrightarrow{\omega}$ come from chiral anomaly, $\mu$ and $\mu_5$ are vector and axial chemical potentials respectively.

It is well known that there is an analogue of chiral anomaly in the condense matter physics \cite{Volovik, Volovik1, Haldane}. In the recent work \cite{SonBerry} this anomaly for the particle number current in the fermi liquid of quasiparticles was discussed. It is due to the presence of Berry curvature term, which modifies the equations of motion for a single quasiparticle, and, therefore affects the kinetics of the system. Provided that there is non-zero flux of Berry curvature through the Fermi surface, the particle current divergence is anomalous. Consequently, this allows for CME and CVE in a usual manner. This was explicitly shown in \cite{SonBerry} for magnetic field, whereas we will generalize this to include rotation and obtain CVE.

For non-zero Berry curvature flux through Fermi surfaces there should be degeneracy points in dispersion relation of system \cite{Volovik1, Volovik3, Haldane}. As discussed in \cite{Volovik3} the dispersion relation in vicinity of degeneracy points is linear if there is no additional symmetry. However, there are known examples when the dispersion relation could be quadratic (for certain directions) in vicinity of degeneracy points. We argue here that the results for CVE for emergent relativistic and non-relativistic spectrum type correspond to each other in the same way as for real relativistic theory. Namely, the linear (relativistic) dispersion yields the full relativistic CVE result, while quadratic (non-relativistic) gives the non-relativistic CVE result (substitution $\mu \rightarrow m$).

It is also noteworthy that in the quantum field theory context there have been various open questions regarding CVE and CME. It wasn't quite clear which particles carry this current, and it was also argued that in some situations it should be prescribed to defects, such as in superfluidity. In the latter case it was observed that the defect picture yields answer which is different from the usual one \cite{Kirilin}. This proves that investigation of the microscopical origin of chiral currents in condensed matter models is indeed important.

\section{Rotating system, Berry curvature and chiral vortical effect}
We follow the approach developed in \cite{SonBerry}. The starting point in the discussion is the equation of motion for a single quasiparticle in a theory with non-zero Berry curvature and in the presence of electromagnetic field \cite{Sundaram}:

\begin{eqnarray}
\dot{\overrightarrow{x}}=\frac{\partial H}{\partial \overrightarrow{p}}-\dot{\overrightarrow{p}}\times \overrightarrow{\Omega}\\
\dot{\overrightarrow{p}}=e(\overrightarrow{E}+\dot{\overrightarrow{x}}\times \overrightarrow{B}),
\end{eqnarray}
where $\Omega$ is Berry curvature. As it was shown in \cite{SonBerry} one can obtain for the particle number density and current the following relation $\partial_t n+ \nabla\cdot j=\frac{k}{4\pi^2}E\cdot B$ which coincides with chiral anomaly in field theory. Here $k$ is quanta of Berry curvature flux through the Fermi surface. Quantization of Berry curvature flux has topological origin (see \cite{Haldane}). It also should be noted that total flux of Berry curvature through all Fermi surfaces is zero \cite{Haldane} so total particle current is conserved and chemical potential could be used for the related charge.  After that consideration one can obtain familiar expression for CME in the system $\overrightarrow{j}_{CME}=\frac{k}{4\pi^2}\mu\overrightarrow{ B}$. Total current is given by sum over all fermi surfaces. If we suppose that there is one surface with $k = +1$ and another with $k = -1$, then the total current vanishes if chemical potentials are equal. However, if they are different, $\mu_{+}  \neq \mu_{-}$ then we obtain:

\begin{eqnarray}
\overrightarrow{j}_{CME}=\frac{\mu_+-\mu_-}{4\pi^2}\overrightarrow{B}.
\end{eqnarray}
That result is very similar to usual answer for CME in field theory if one replaces $\frac{\mu_+-\mu_-}{2}\rightarrow\mu_5$, so electrons on two different Fermi surfaces could be treated as analogous of two chiralities in field theory.

To consider CVE we should include the possibility of rotation. In general case arbitrary dispersion relation appears to be linear in vicinity of degeneracy points. According to \cite{Zakharov} one could take rotation into account for effectively chiral fermions by substitution $A_i\rightarrow A_i+\mu v_i$, where $v_i$ is the velocity of liquid. In the case of uniform rotation using consideration of \cite{SonBerry} one gets for CVE (\ref{chiral}), where as mentioned above $\mu_5=\frac{\mu_+-\mu_-}{2}$. However there are  systems which exhibit non-linear spectrum (in certain directions) in vicinity of degeneracy points (see \cite{Volovik3, Volovik4}). In particular lets consider quadratic spectrum in xy-plane. If such system is rotating around z-axis then one should take non-relativistic limit of (\ref{chiral}). We will use vector notation below for convenience; however, we will only consider $\omega$ along z axis. Consideration of CVE as result of Larmor's theorem was proposed in \cite{KharzeevZhitnitsky}. It should be emphasized that we will consider only uniform rotation. This means that the liquid rotates as a whole, so that the Fermi surface is formed by momentum with respect to rotating coordinate system. This means that the motion of quasiparticles should be described with respect to the rotating frame of reference. In particular, quasiparticles will "feel" inertial forces such as Coriolis force. We will neglect centrifugal force thereafter, and hence the angular velocity is supposed to be small enough to validate that assumption. The equations of the motion in the rotating frame then take form:

\begin{eqnarray}
\dot{\overrightarrow{x}}=\frac{\partial H}{\partial \overrightarrow{p}}-\dot{\overrightarrow{p}}\times\overrightarrow{\Omega}\\
\dot{\overrightarrow{p}}= 2 m \dot{\overrightarrow{x}}\times \overrightarrow{\omega},
\end{eqnarray}
where $\omega$ is the angular velocity, $\Omega$ is  Berry curvature and $m$ is the effective mass of the quasiparticle.  Following the \cite{SonBerry} the action for single particle in a theory with Berry curvature could be constructed and one gets

\begin{eqnarray}
S=\int \left(p\cdot\dot x+  m [\omega\times x]\cdot\dot x-\mathcal A(p)\cdot\dot p-H(p,x)\right)dt,
\end{eqnarray}
where $\mathcal A_i(p)$ is fictitious vector-potential for Berry curvature in momentum space,  $\Omega_i=\epsilon_{ijk}\partial_j \mathcal A_k$ and form of "Coriolis" term is appropriate for quadratic spectrum (and for emergent relativity by substituting $m\rightarrow\mu$). Combining $x_i$ and $p_i$ in one set of variables $\xi_a$, where $a=1..6$ the action becomes $S=\int \left(-\omega_a \dot\xi_a-H(\xi)\right)dt$. Equations of motion following from this action can be written as          $\dot\xi_a=-\{\xi_a,\xi_b\}\frac{\partial H}{\partial \xi_b}$. As it was shown in \cite{Duval} Poison brackets for this action are

\begin{eqnarray}
\{x_i,x_j\}=\frac{\epsilon_{ijk}\Omega_k}{1+2m\omega\cdot \Omega}\\
\{p_i,p_j\}=-\frac{2m\epsilon_{ijk}\omega_k}{1+2m\omega\cdot \Omega}\\
\{p_i,x_j\}=\frac{\delta_{ij}+2m\Omega_i\omega_j}{1+2m\omega\cdot \Omega},
\end{eqnarray}
and invariant measure of phase space is $d\Gamma=(1+2m\omega\cdot \Omega)\frac{d^3pd^3x}{(2\pi)^3}$. This modification of Poisson bracket (in comparison with the conventional $ \{x_i,x_j\} = 0, \{p_i,p_j\}= 0, \{p_i, x_j\} = \delta_{ij}$) alters the kinetics of the liquid (see \cite {SonBerry} for details).

In particular, the current takes form:

\begin{eqnarray}
\label{current}
\overrightarrow{j}=\int \frac{d^3p}{(2\pi)^3}\left(-\epsilon_p\frac{\partial n_p}{\partial \overrightarrow{p}}-2m\epsilon_p\left(\overrightarrow{\Omega}\cdot\frac{\partial n_p}{\partial \overrightarrow{p}}\right)\overrightarrow{\omega}-\epsilon_p[\overrightarrow{\Omega}\times \frac{\partial n_p}{\partial \overrightarrow{x}}]\right).
\end{eqnarray}
Here $n_p(x)$ is the distribution function for the quasiparticles and $\epsilon_p$ is the energy of a given quasiparticle with momentum $p$. The first term in the bracket is just $n_p\overrightarrow{v}$, where $\overrightarrow{v}=\frac{\partial \epsilon_p}{\partial \overrightarrow{p}}$, and naturally, it vanishes in thermal equilibrium. The third term is also zero in ground state because it involves spatial derivatives. To consider the second term in (\ref{current}) we use Fermi distribution $n_p=\left(\exp{\left(\frac{\epsilon_p-\mu}{T}\right)}+1\right)^{-1}$, where $\mu$ is chemical potential (Fermi energy). After simple transformation we obtain $\overrightarrow{j}_{CVE}=\frac{k\mu m}{2\pi^2}\overrightarrow{\omega}$, $k=\frac{1}{2\pi}\int \Omega\cdot dS$ and if for instance $k=\pm2$ (for quadratic spectrum) then the sum over surfaces is:

\begin{eqnarray}
\label{result}
\overrightarrow{j}_{CVE}=|k|\frac{\mu_5 m}{\pi^2}\overrightarrow{\omega},
\end{eqnarray}
where $\mu_5 = \frac{\mu_+ - \mu_-}{2}$.

For arbitrary dispersion relation and behavior near degeneracy points the difference in chemical potentials of Fermi surfaces could be made by applying both electric and magnetic fields for finite time. One can then argue the existence of an analogue of a chiral battery, proposed in \cite{Kharzeev}. The only difference is that the current might be induced by rotating the sample, rather than by applying magnetic field.

\section{Conclusion}
We show that in rotating Fermi liquid with nonzero flux of Berry curvature there appears an effect similar to CVE.  For quadratic dispersion in vicinity of degeneracy points it could be readily obtained by replacing $B\rightarrow 2m\omega$ in the result for CME in \cite{SonBerry} according to quantum analogy of Larmor theorem.  It corresponds to the non-relativistic limit of well-known result (\ref{chiral}) since in the said limit one can replace $\mu \rightarrow m$. It seems very natural and intuitive that the answer for the effectively non-relativistic spectrum (quadratic) appears as a non-relativistic limit of the effectively relativistic case (linear dispersion relation).
One should note that at the first glance the result (\ref{result}) for CVE depends on renormalization picture because of quasiparticle effective mass.  But it conforms with usual result if one treats mass as external parameter of effective theory in analogy with chemical potentials. It should be noted that  chemical potentials can be treated as result of strong interaction, responsible for forming liquid and in that sense similar to effective mass of quasiparticle in non-relativistic limit as mentioned above. Other reasoning for the identification of effective mass as the non-relativistic limit of the chemical potential, from the point of view of hydrodynamics, is given in \cite{Teaney}.

We would like to point out that our treatment describes the effect in the leading order in angular velocity. The question how is the result modified when we take into account higher corrections in  $\Omega$ is not quite clear and requires further research. Some approaches to the problem are given in \cite{Landsteiner, Vilenkin}.

We also emphasize that in condensed matter context we have great experimental and theoretical grounds to test origins of the chiral currents: whether they are prescribed to defects, which particles (or quasiparticles) carry them etc.

\section{Aknowledgments}
We would like to acknowledge discussions with A.Krikun, S. Guts, V. Zakharov and M. Zubkov. Authors are thankful to G.E. Volovik for valuable comments. The work was supported by Grant RFBR-11-02-01227-a and by the Federal
Special-Purpose Program "Cadres" of the Russian Ministry of Science and Education (contracts 02.740.11.0571, 8174 and 8376). The work was also partly supported by the Dynasty foundation and FAIR program for master students. The work of V.P.K. was also supported by DAAD Leonhard-Euler-Stipendium 2011-2012.

\end{document}